\documentclass[12pt]{article}
\usepackage{epsfig,amsfonts,amssymb}
\usepackage{hyperref}
\usepackage{cite}
\input epsf.sty
\def\ZZZ{{\hbox{ Z\kern-1.6mm Z}}}
\def\RRR{{\hbox{ R\kern-2.4mm R}}}
\def\CCC{{\hbox{ C\kern-2.0mm C}}}
\def\zzz{{\hbox{z\kern-1mm z}}}

\newcommand{\qeq}{{\hbox{=\kern-2.3mm ? \kern.5mm }}}
\renewcommand{\qeq}{=}

\newcommand{\be}{\begin{equation}}
\newcommand{\ee}{\end{equation}}
\newcommand{\ben}{\begin{eqnarray}\displaystyle}
\newcommand{\een}{\end{eqnarray}}

\newcommand{\bea}[1]{\begin{eqnarray}\label{#1} }
\newcommand{\eea}{\end{eqnarray}}

\newcommand{\refb}[1]{(\ref{#1})}
\newcommand{\p}{\partial}

\def\one{{\hbox{ 1\kern-.8mm l}}}
\def\zero{{\hbox{ 0\kern-1.5mm 0}}}

\usepackage{cite}

\begin{document}

\baselineskip 24pt

\begin{center}
{\Large \bf Interpreting the M2-brane Action
}

\end{center}

\vskip .6cm
\medskip

\vspace*{4.0ex}

\baselineskip=18pt

\centerline{\large \rm   Shamik Banerjee and 
Ashoke Sen }

\vspace*{4.0ex}

\centerline{\large \it Harish-Chandra Research Institute}

\centerline{\large \it  Chhatnag Road, Jhusi,
Allahabad 211019, INDIA}

\vspace*{1.0ex}
\centerline{E-mail:  bshamik@mri.ernet.in, 
sen@mri.ernet.in, ashokesen1999@gmail.com}

\vspace*{5.0ex}

\centerline{\bf Abstract} \bigskip

The world-volume theory of multiple M2-branes proposed recently
has a free scalar field. For large vev of this
scalar field the world-volume action reduces to that of 
multiple D2-branes with Yang-Mills coupling proportional to the
vev. We suggest that the correct interpretation of this scalar field
is as the radial position of the M2-brane center of mass in
a cylindrical polar coordinate system. Regarding the azimuthal
angle as compact we can regard this as a set of coincident D2-branes
in type IIA string theory with varying dilaton and metric.
We find that the effective world-volume theory on the D2-branes
has Yang-Mills coupling proportional to the radial coordinate;
furthermore the radial coordinate satisfies free field equations of
motion. This agrees with the corresponding results derived from
the M2-brane world-volume theory.

\vfill \eject

\baselineskip=18pt

Since the discovery of a general class of $(2+1)$ dimensional
superconformal field theories due to Bagger, Lambert and
Gustavson\cite{0611108,0709.1260,0711.0955,0712.3738,0802.3456}, 
following earlier work
of  \cite{0411077,0412310}, there has been 
much activity in trying to 
interpret
these theories as the world-volume action of multiple 
M2-branes, as well as exploring various other aspects
of these theories\cite{0803.3218}-\cite{0805.3662}.
The simplest theory based on an $SO(4)$ algebra turned out
to describe  a theory of M2-branes
on an orbifold\cite{0804.1114,0804.1256}. 
Furthermore, this theory has a novel
Higgs mechanism whereby by giving vev to a scalar field on
the world-volume we can recover the D2-brane action with
an effective Yang-Mills coupling proportional to the 
vev\cite{0803.3218}. While
this result looked puzzling at the first sight, a simple
interpretation was found in \cite{0804.1256}, 
where it was shown that in
appropriate region of the M2-brane moduli space the transverse
space of the 
M2-branes has a compact direction
whose radius varies slowly as a function of one of the
non-compact
coordinates. 
By exploiting the duality between M-theory on a
circle and type IIA string theory\cite{9501068,9503124}, 
we can locally regard this theory
as a type IIA string theory with slowly varying background and the
M2-branes as D2-branes moving in this background. 
This provides a natural explanation of how the vev of a
world-volume scalar field could appear as a coupling constant of
the D2-brane world-volume theory.

More recently refs.\cite{0805.1012,0805.1087,0805.1202} 
proposed a theory of multiple M2-branes
in flat space-time. The theory uses an indefinite metric in order
to evade a no-go theorem\cite{0804.2662,0804.3078}, 
and much is yet to be understood
about this theory. However one of the puzzling features of this
theory is that it has a free scalar field, and giving a large vev
to this scalar again produces the world-volume theory of
multiple D2-branes whose Yang-Mills coupling is proportional
to the scalar vev. Our goal in this note is to try to give an
interpretation of this phenomenon along the lines of
\cite{0804.1256}. However our analysis will have much less control;
unlike in the analysis of \cite{0804.1256} 
where one could take an
appropriate limit involving the order of the orbifold group
and the scalar vev to keep the effective Yang-Mills coupling
on the D2-branes small, we do not have such a control parameter.
This however is to be expected if we are to describe M2-branes
in flat eleven dimensional space-time where we have no
adjustable coupling constant.

In a nutshell our proposal is the following. We split the 
eight transverse
coordinates of the M2-brane into two sets, -- two described
using polar coordinate system and the rest of the six
described using the Cartesian coordinates. Away from the
origin of the polar coordinates, we can regard the azimuthal
angle as a compact coordinate and regard the M2-branes as  D2-branes
moving in a type IIA background whose dilaton and the metric
vary as a function of the radial coordinate. A naive classical
analysis shows that the effective Yang-Mills
coupling constant on the
D2-branes is proportional to the radial coordinate, thereby
suggesting that we interpret the radial coordinate as the
scalar field in the M2-brane world-volume theory whose vev
controls the D2-brane coupling. Furthermore we can compute the
equation of motion of the radial coordinate by regarding it
as a scalar field on the D2-brane world-volume and find that
it satisfies free field equations of motion. This agrees with the
corresponding result derived from the M2-brane 
world-volume action.

We shall now proceed to describe our analysis.
Let $x^\mu$ ($0\le \mu\le 2$) be the world-volume coordinates
of the M2-brane, $(r,\theta)$ denote the polar coordinates of
two of the transverse directions of the M2-brane and 
$y^m$ ($3\le m\le 8$)
be the rest of the transverse directions. In this 
coordinate system
the eleven dimensional
flat metric is given by
\be \label{e1}
ds_{11}^2 = \eta_{\mu\nu} dx^\mu dx^\nu + (dr^2 + r^2 d\theta^2)
+ dy^m dy^m\, .
\ee
We shall now interprete the coordinate $\theta$ as the compact
direction and regard the theory as type IIA string theory in a
space-time labelled by $(\vec x, r, \vec y)$. Then by standard
rules\cite{9503124} the type IIA dilaton
$\phi$ and the type IIA metric $ds_{IIA}^2$ are given by
\be \label{e2}
e^{2\phi} = r^3, \qquad ds_{IIA}^2\equiv (g_{\mu\nu}dx^\mu
dx^\nu + G_{rr} dr^2 + G_{mn} dy^m dy^n) =
r ( 
\eta_{\mu\nu} dx^\mu dx^\nu + dr^2 + dy^m dy^m)\, .
\ee
In this picture $N$ coincident M2-branes placed at 
$(r,\theta,\vec y)$
can be regarded as $N$ coincident D2-branes placed
at $(r,\vec y)$. The part of the D2-brane
world-volume action containing
the $SU(N)$ gauge fields will be given by
\be \label{e3}
-{1\over 4} \, \int d^3 x \,
e^{-\phi} \sqrt{-\det g} \, g^{\mu\rho}
\, g^{\nu\sigma} \, Tr\left[F_{\mu\nu} F_{\rho\sigma}\right]
= -{1\over 4} \, \int d^3 x \, r^{-2} \, \eta^{\mu\rho}
\, \eta^{\nu\sigma} \, 
Tr\left[F_{\mu\nu} F_{\rho\sigma}\right]\, .
\ee
{}From this we see that the effective Yang-Mills coupling $g_{YM}$
is given by
\be \label{e4}
g_{YM} = r\, .
\ee
On the other hand the world-volume scalar field $R$ on the M2-brane,
labelling the centre of mass $r$ coordinate, has an action proportional
to
\be \label{e5}
- {1\over 2}\, 
\int d^3 x\, \left[ e^{-\phi} \sqrt{-\det g} \, g^{\mu\nu} G_{rr}
\right]_{r=R} 
\p_\mu R
\p_\nu R = - {1\over 2}\,
\int d^3 x \, \eta^{\mu\nu} \, \p_\mu R \p_\nu R\, .
\ee
Thus the world-volume field
$R$ satisfies the free field equation of motion
\be \label{e6}
\eta^{\mu\nu} \p_\mu \p_\nu R = 0\, ,
\ee
and the vev of this field determines the $r$-coordinate of the
branes.

Let us compare this with the results of 
\cite{0805.1012,0805.1087,0805.1202}. 
For definiteness we shall
use the convention of \cite{0805.1012}. In this paper the 
membrane action contains
a scalar field $X^{-(8)}$ whose vacuum expectation value controls
the strength of the Yang-Mills coupling. 
Thus we can identify it with the scalar field $R$ introduced above.
Furthermore the field $X^{-(8)}$ is known to satisfy the equations
of motion of a free field:
\be \label{e77}
\eta^{\mu\nu}\p_\mu\p_\nu X^{-(8)}=0\, .
\ee
This agrees with \refb{e6},
providing a non-trivial consistency check on our 
proposal.\footnote{Note that for large $r$, \i.e.\ large
$X^{-(8)}$ where the M2-brane theory reduces to the D2-brane
theory, the string coupling is large but the curvature and the
dilaton gradient measured in the string scale is small.}

If this proposal is correct then this implies that the apparent 
dependence of the M2-brane world-volume
theory on the vacuum expectation value of the
field $X^{-(8)}$ is fake; different vacuum expectation values
of $X^{-(8)}$ correspond to placing the membrane at different
values of the radial coordinate $r$. It will be interesting to see
if this symmetry is present in the theories proposed in
\cite{0805.1012,0805.1087}; 
this will involve a non-trivial transformation on the
fields since from the geometric viewpoint described above
this will amount to transforming from polar coordinates
around a given origin to the polar coordinates around a different
origin. It is of course possible that the symmetry is not visible in
the classical theory, and only after quantization we shall
discover this symmetry. This also suggests that there may be a
simpler version of the M2-brane world-volume theory that uses the
Cartesian coordinate system instead of cylindrical polar cordinates
for describing the transverse space.

\end{document}